\documentclass[11pt,a4paper]{article}

\usepackage[T1]{fontenc}
\usepackage[utf8]{inputenc}
\usepackage{lmodern}
\usepackage{geometry}
\usepackage{amsmath,amssymb,amsthm}
\usepackage{enumitem}
\usepackage[hidelinks]{hyperref}

\title{Comment on Classical-Gravity--Quantum-Matter Claims About Gravity-Mediated Entanglement}
\author{
\large Mikołaj Sienicki\thanks{Polish-Japanese Academy of Information Technology, ul. Koszykowa 86, 02-008 Warsaw, Poland, European Union.} 
\quad and \quad
Krzysztof Sienicki\thanks{Chair of Theoretical Physics of Naturally Intelligent Systems (NIS\textsuperscript{\textcopyright}), Lipowa 2/Topolowa 19, 05-807 Podkowa Leśna, Poland, European Union.}
}
\date{\today}

\begin{document}
\maketitle

\begin{abstract}
A recent paper by Aziz and Howl (Nature 2025) argues that, when quantum matter is described at the level of quantum field theory and coupled to a classical gravitational field, higher--order processes can generate entanglement between two spatially separated masses. A contemporaneous critical note (Marletto, Oppenheim, Vedral, Wilson, arXiv:2511.07348v1) shows that, in the actual nonrelativistic limit used in that analysis, the interaction becomes ultra-local, the total unitary factorizes, and no entanglement is produced from a product input. In this comment we (i) restate the core point of that critique, (ii) give a channel-theoretic reformulation that makes the conclusion model-independent, and (iii) clarify the distinction between \emph{activation} of entanglement in already-quantum matter and genuine \emph{mediation} of entanglement by a classical field. With these clarifications in place, the standard BMV inference---that the observation of gravity-mediated entanglement strongly indicates nonclassical gravitational degrees of freedom---remains intact.
\end{abstract}

\textbf{Keywords:}
quantum gravity; gravity-mediated entanglement; classical gravity; semiclassical gravity;
BMV proposals; quantum channels; entanglement activation; hybrid quantum--classical dynamics.

\section{Introduction}
There is now an active research programme, initiated by Bose \emph{et al.}~\cite{Bose2017} and Marletto--Vedral~\cite{MarlettoVedral2017}, which proposes to test the \emph{quantum} nature of gravity by observing whether two massive systems, coupled only via gravity, can become entangled. The usual logical template can be phrased as follows:
\begin{quote}
If two quantum systems become entangled via some mediator, then that mediator must itself have nonclassical degrees of freedom (or, at the very least, implement dynamics beyond LOCC).
\end{quote}
This basic idea has been analyzed, sharpened and generalized in a number of follow-up works. What these analyses share is the conclusion that a \emph{purely classical} mediator, restricted to classical communication and classical control, cannot create entanglement between initially separable quantum systems.

Aziz and Howl (AH) \cite{AzizHowl2025} recently questioned how strong this inference really is. They suggest that, once the matter sector is treated as fully quantum (at the level of QFT) while the gravitational field remains classical, higher--order processes in the matter sector can mimic gravity-mediated entanglement. Almost immediately, Marletto, Oppenheim, Vedral and Wilson (MOVW) showed in \cite{MOVW2025} that, in the concrete nonrelativistic limit actually used by AH, the model does not in fact generate entanglement. 

In what follows we give a short, largely self-contained summary of both papers and then add two observations. First, we recast the discussion in the language of quantum channels, where the limitations of a classical mediator become especially transparent. Second, we explicitly separate the notion of \emph{entanglement activation} in already-quantum matter from that of \emph{entanglement mediation} by gravity itself. Taken together, these points support the view that BMV-type experiments continue to provide a clean operational test for nonclassicality in gravity.

\section{Summary of Aziz and Howl}
The starting point of the AH paper is the observation that many previous arguments against \emph{classical} gravity mediating entanglement assume a relatively simple quantum-mechanical description of the two masses (as few-level systems or wave packets) together with an effectively Newtonian gravitational potential. AH argue that this treatment of matter is too coarse. If the matter sector is genuinely quantum field theoretic, then, even when the gravitational field is taken to be classical and coupled in a semiclassical, mean-field fashion to the expectation value of the stress-energy tensor, one can write down fourth-order perturbative processes in which virtual matter excitations propagate between the two massive systems. AH interpret the resulting contribution to the effective interaction as an entangling term.

Two features of the AH analysis are worth emphasizing:
\begin{enumerate}[label=(\alph*)]
  \item The candidate entangling term appears only at \emph{fourth} order in the perturbative expansion, in contrast to the standard quantum-gravity picture where a second-order exchange already suffices to produce entanglement.
  \item The coupling of gravity to matter is of the semiclassical type: the gravitational field responds to $\langle \hat{T}_{\mu\nu} \rangle$, while the matter degrees of freedom are treated fully quantum mechanically.
\end{enumerate}
On this basis AH suggest that, even if an experiment observed entanglement between two test masses, this would not by itself force us to conclude that gravity must be quantized.

\section{Summary of the critique (arXiv:2511.07348v1)}
The MOVW critique \cite{MOVW2025} revisits the AH derivation in the precise nonrelativistic limit that AH actually adopt. They identify three main problems.

\subsection{Ultra-locality and factorization}
In the relevant limit, the effective Hamiltonian for the two separated masses reduces to a sum of strictly local terms,
\begin{equation}
H \simeq H_A \otimes \mathbb{I}_B + \mathbb{I}_A \otimes H_B.
\end{equation}
The corresponding time evolution then factorizes,
\begin{equation}
U(t) = e^{-i H t} = e^{-i H_A t} \otimes e^{-i H_B t} = U_A(t) \otimes U_B(t),
\end{equation}
so the dynamics is generated by a product unitary. Such a unitary cannot turn an initially separable state into an entangled one. Therefore, under the very approximations that AH themselves impose, the model does not generate entanglement.

\subsection{Inconsistent handling of spatial/gradient terms}
MOVW also observe that AH effectively \emph{discard} gradient terms of the form $(\nabla \hat{\phi})^2$ when simplifying the interaction, but later, in the evaluation of the propagator, implicitly rely on nonlocal features that such gradient terms would normally provide. In a consistent nonrelativistic treatment, the propagator between non-overlapping wave packets vanishes, and with it the cross-term that AH identify as the source of entanglement. Once the approximation is implemented consistently, the would-be entangling contribution disappears.

\subsection{Perturbative artefacts}
Finally, MOVW emphasize a more general point: truncating the series expansion of a \emph{product} unitary can give intermediate terms that look entangling, even though the exact unitary is not. Since the AH effect only appears at fourth order, it is natural to suspect that it is precisely such a truncation artefact. In other words, the entangling-looking contribution in the truncated series cancels out once the full product structure of the unitary is taken into account.

\section{Channel-theoretic reformulation}
A convenient way to make the critique less dependent on the detailed field-theoretic model is to express it in terms of quantum channels. Suppose that the gravitational field is genuinely classical, in the sense that it can at most supply a classical label $c$ (representing, for example, the value of the classical field or the outcome of some effective classical procedure) on which each subsystem conditions its local evolution. The most general two-party channel of this kind is
\begin{equation}
\label{eq:classical-channel}
\mathcal{E}(\rho_{AB}) = \sum_{c} p(c) \bigl(U^{(c)}_A \otimes U^{(c)}_B\bigr)\, \rho_{AB}\, \bigl(U^{(c)}_A \otimes U^{(c)}_B\bigr)^\dagger.
\end{equation}
Channels of the form \eqref{eq:classical-channel} are separable: they are convex mixtures of local unitaries, and therefore cannot generate entanglement from an initially product state.

This leads to a simple diagnostic. If a purported ``classical-gravity'' model is intended to be of the form \eqref{eq:classical-channel}, but a perturbative calculation seems to show that it creates entanglement, then one of two things must be happening:
\begin{enumerate}[label=(\roman*)]
  \item the gravitational sector has, perhaps implicitly, been allowed to carry quantum (non-classical) correlations, so that the model is no longer genuinely classical; or
  \item the perturbative truncation has introduced spurious cross-terms that would cancel in the full (non-truncated) series.
\end{enumerate}
This is precisely the structure that MOVW identify in the AH derivation: what looks like an entangling contribution at fourth order is an artefact of the approximations used, not a generic feature of classical gravity.

\section{Activation vs.\ mediation}
Even if one were to take the AH term at face value and assume it survived all consistency checks, its physical interpretation would still need to be more modest. In the AH framework, the matter sector is fully quantum and, by itself, capable of generating entanglement. The classical gravitational variable merely selects which quantum evolution is implemented. It acts as a classical control knob for an already-quantum machine.

It is therefore more accurate to describe the situation as \emph{entanglement activation} in the matter sector, driven by classical control, rather than as genuine \emph{entanglement mediation} by a classical gravitational field. Put differently, the nonclassical resource responsible for the entanglement resides squarely in the matter degrees of freedom, not in the gravitational ones.

\section{Relation to hybrid quantum--classical gravity models}
There is a broader literature on hybrid quantum--classical models of gravity (for example, semiclassical gravity with stochastic corrections, or measurement-and-feedback schemes) that is concerned with maintaining positivity and avoiding superluminal signalling. A recurring outcome of these constructions is that a purely classical mediator, obeying these consistency conditions, does \emph{not} generate entanglement between two separated quantum systems.

The AH effect appears precisely when one relaxes, or applies non-uniformly, the kinds of constraints that such hybrid models typically impose. From this perspective, the AH contribution looks like a feature of particular modelling choices rather than a robust prediction of the general framework ``classical gravity + quantum matter''. This reinforces the idea that BMV-type experiments remain conceptually clean as probes of gravity's nonclassicality.

\section{Implications for BMV-type experiments}
The original motivation of BMV-type proposals \cite{Bose2017,MarlettoVedral2017} was to provide an \emph{operational} witness of the nonclassical character of gravity: if two masses become entangled and the only interaction between them is gravitational, then the mediator responsible for this entanglement must be nonclassical. The AH paper was naturally read as a challenge to this logic, by suggesting that entanglement might arise even with a classical gravitational field.

The analysis of MOVW, together with the channel-theoretic argument presented above, restores the original inference. For a genuinely classical gravitational mediator, the relevant evolution is of the separable form \eqref{eq:classical-channel} and cannot create entanglement from a product input. Therefore, if a BMV-type experiment were to observe entanglement where gravity is the only plausible mediator, this would still count as strong evidence that gravitational degrees of freedom are nonclassical.

\section{Conclusion}
Aziz and Howl have usefully highlighted that combining classical gravity with fully quantum matter is subtle, and that higher--order matter processes can be easily misinterpreted. However, in the specific limit they actually work with, their model does not provide an example of gravity-mediated entanglement: the interaction becomes ultra-local and the evolution factorizes into independent dynamics on each mass. When the problem is phrased in terms of separable channels and when one carefully distinguishes entanglement activation from genuine mediation, this conclusion becomes transparent and largely independent of model-specific details.

On this reading, the central claim underlying entanglement-based tests of the quantum nature of gravity remains well supported: if an experiment observes entanglement between two masses that interact only via gravity, then the mediator responsible cannot be purely classical.

\paragraph*{Note added in proof.}
After this comment was completed, Schneider, Huggett and Linnemann
provided an independent analysis of gravitationally induced
entanglement experiments in the Newton--Cartan formulation of the
Newtonian limit of general relativity \cite{SchneiderHuggettLinnemann2025}.
They show that, when classical gravity is treated as a single
Newton--Cartan connection, gravitational mediation by itself cannot
generate entanglement between spatially separated masses: the phase
that would be responsible for entanglement vanishes when the two
branches of the experiment share the same classical geometry. This
geometric result is fully consistent with, and complementary to, the
channel-theoretic argument of Sec.~4 above, according to which a
genuinely classical gravitational mediator implements a separable
evolution and therefore cannot create entanglement from a product
input state.

\end{document}